\documentclass[prl,aps,floatfix,superscriptaddress,nofootinbib,twocolumn]{revtex4-1}

\usepackage{amssymb}
\usepackage{amsmath}
\usepackage{graphicx}
\usepackage[usenames,dvipsnames]{xcolor}
\usepackage{hyperref}
\usepackage[all]{hypcap} 
\hypersetup{colorlinks=true,
	citecolor=NavyBlue,
	linkcolor=NavyBlue,
    	filecolor=NavyBlue,
	urlcolor=NavyBlue,
   }
\def\ba{\begin{eqnarray}}
\def\ea{\end{eqnarray}}

\begin{document}  

\begin{flushleft}      
KCL-PH-TH-2021-25\\
CERN-TH-2021-061
\end{flushleft}
\begin{flushleft}      
\end{flushleft}

\title{First Constraints on Nuclear Coupling of Axionlike Particles from the Binary Neutron Star Gravitational Wave Event GW170817}

\author{Jun Zhang}
\email{jun.zhang@imperial.ac.uk}
\affiliation{Theoretical Physics, Blackett Laboratory, Imperial College, London, SW7 2AZ, UK}
\affiliation{Illinois Center for Advanced Studies of the Universe \&
Department of Physics, University of Illinois at Urbana-Champaign,
Urbana, Illinois 61801, USA}
\author{Zhenwei Lyu}
\email{zlyu@uoguelph.ca}
\affiliation{University of Guelph, Guelph, Ontario N2L 3G1, Canada}
\affiliation{Perimeter Institute for Theoretical Physics, Waterloo, Ontario N2L 2Y5, Canada}
\author{Junwu Huang}
\email{jhuang@perimeterinstitute.ca}
\affiliation{Perimeter Institute for Theoretical Physics, Waterloo, Ontario N2L 2Y5, Canada}
\author{Matthew C. Johnson}
\email{mjohnson@perimeterinstitute.ca}
\affiliation{Department of Physics and Astronomy, York University, Toronto, Ontario, M3J 1P3, Canada}
\affiliation{Perimeter Institute for Theoretical Physics, Waterloo, Ontario N2L 2Y5, Canada}
\author{Laura Sagunski}
\email{sagunski@itp.uni-frankfurt.de}
\affiliation{Institute for Theoretical Physics, Goethe University, 60438 Frankfurt am Main, Germany}
\author{Mairi Sakellariadou}
\email{mairi.sakellariadou@kcl.ac.uk}
\affiliation{Theoretical Particle Physics and Cosmology Group, Physics Department, King's College London, University of London, Strand, London WC2R 2LS, UK}
\affiliation{Theoretical Physics Department, CERN, Geneva, Switzerland}
\author{Huan Yang}
\email{hyang@perimeterinstitute.ca}
\affiliation{University of Guelph, Guelph, Ontario N2L 3G1, Canada}
\affiliation{Perimeter Institute for Theoretical Physics, Waterloo, Ontario N2L 2Y5, Canada}

\begin{abstract} 
Light axion fields, if they exist, can be sourced by neutron stars due to their coupling to nuclear matter, and play a role in binary neutron star mergers. We report on a search for such axions by analyzing the gravitational waves from the binary neutron star inspiral GW170817. We find no evidence of axions in the sampled parameter space. The null result allows us to impose constraints on axions with masses below $10^{-11} {\rm eV}$ by excluding the ones with decay constants ranging from $1.6\times10^{16} {\rm GeV}$ to $10^{18} {\rm GeV}$ at a $3\sigma$ confidence level. Our analysis provides the first constraints on axions from neutron star inspirals, and rules out a large region in parameter space that has not been probed by the existing experiments.
\end{abstract}

\maketitle 

\emph{Introduction.---}%
Axions are hypothetical scalar particles that generally appear in many fundamental theories. An archetypal example is the QCD axion, a pseudoscalar field proposed to solve the strong CP problem \cite{Peccei:1977hh,Peccei:1977ur,Weinberg:1977ma,Wilczek:1977pj}. Light axions are also a unique prediction of string theory~\cite{Svrcek:2006yi, Arvanitaki:2009fg}, as well as one of the most compelling candidates for dark matter~\cite{Preskill:1982cy,Abbott:1982af,Dine:1982ah}.

Axions have been constrained by measuring the energy loss and energy transport in various astrophysical objects, such as stars \cite{Raffelt:1985nk,Raffelt:1990yz,Anastassopoulos:2017ftl} and supernova 1987A \cite{Raffelt:1987yt,Chang:2018rso}. Further constraints can be imposed if axions make up all of the dark matter in our universe \cite{Sikivie:1983ip,Blum:2014vsa, Abel:2017rtm,Du:2018uak,Sibiryakov:2020eir}. 
In addition, axions with weak self-interactions could lead to black hole superradiance, and hence are constrained by the black hole spin measurements \cite{Arvanitaki:2014wva,Gruzinov:2016hcq,Davoudiasl:2019nlo,Stott:2020gjj,Ng:2020ruv,Baryakhtar:2020gao}, the polarimetric observations \cite{Chen:2019fsq}, and the gravitational waves (GWs) emitted by the superradiance cloud \cite{Arvanitaki:2016qwi,Zhu:2020tht,Brito:2017wnc,Brito:2017zvb,Tsukada:2018mbp,Palomba:2019vxe,Sun:2019mqb}. Bosonic fields may also form compact objects that have GW implications \cite{CalderonBustillo:2020srq}.

\begin{figure}[tp]
\centering 
\includegraphics[height=0.34\textwidth]{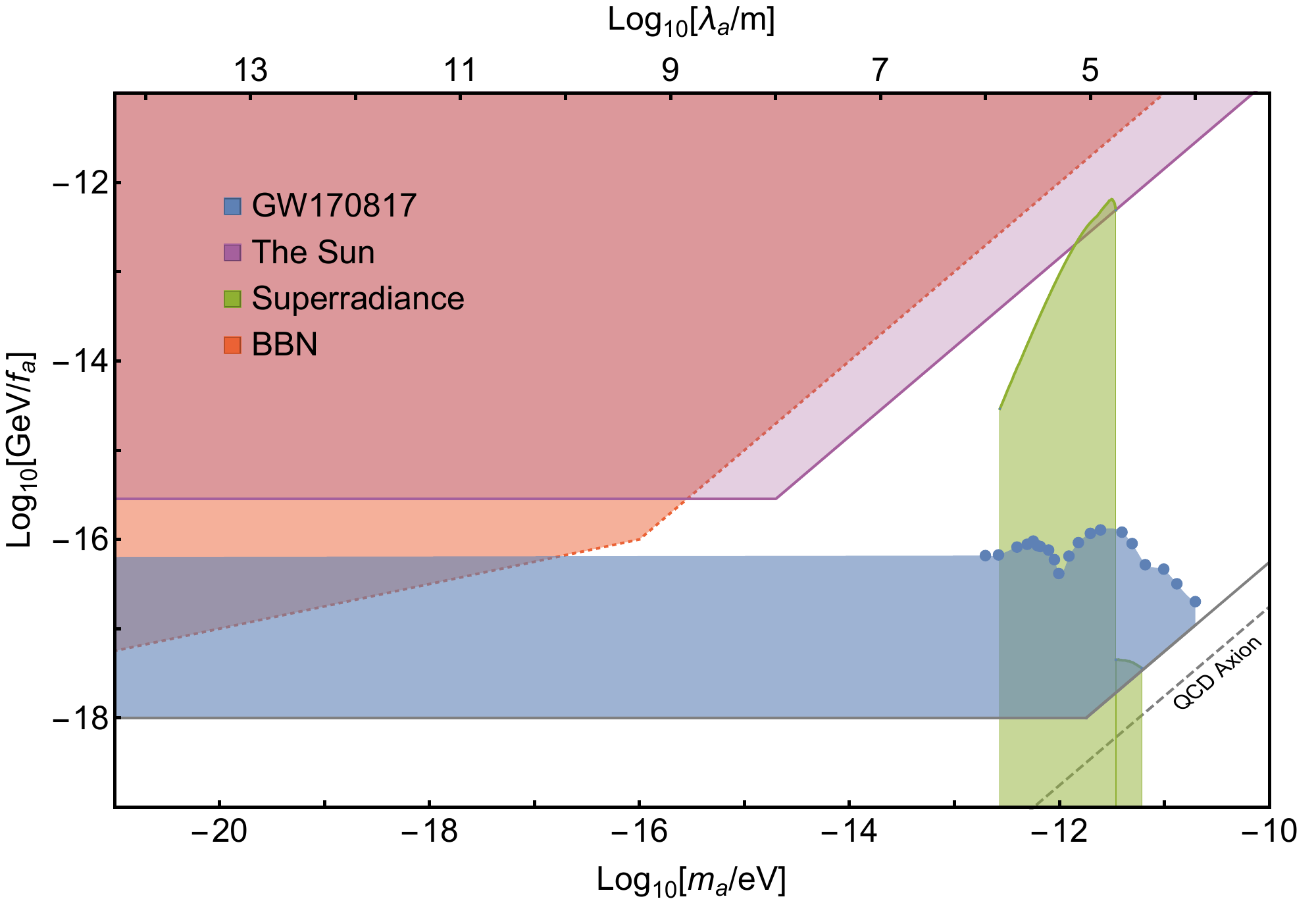} 
\caption{Constraints on the axion parameter space. $m_a$ is the mass of the axion field and $f_a$ is the axion decay constant. The blue dots show the masses of axion fields that are sampled in this Letter, and the corresponding $3\sigma$ constraints on $f_a$ from GW170817. For $f_a \gtrsim 1.6\times10^{16} \rm{GeV}$, the GW170817 data cannot distinguish waveforms with $m_a \lesssim {\rm10^{-13} eV}$, allowing us to extrapolate the constraints on small $m_a$ to the massless limit and to exclude the existence of axions in the blue regime. Axions in the purple region could also be significantly sourced by the Earth and the Sun, and hence are excluded~\cite{Hook:2017psm}. In addition, we show the $3\sigma$ constraint from the spin measurements of the stellar mass black holes (Cyg X-1 and GRS 1915+105) ~\cite{Baryakhtar:2020gao} (in green), as well as the $1\sigma$ constraint on axion dark matter from the BBN~\cite{Blum:2014vsa} (in red). 
}\label{fig:cons}
\end{figure}

In this Letter, we report on a new search for certain axions using GW170817, the GWs from a binary neutron star (NS) inspiral detected by LIGO and Virgo~\cite{TheLIGOScientific:2017qsa}\footnote{Although the scenario of GW170817 being a neutron star-black hole merger can not be ruled out, the astrophysical processes to produce a black hole with neutron-star mass are generally considered to be exotic \cite{Yang:2017gfb}.}. We focus on axions that couple to nuclear matter in the same way as the QCD axion, but with masses that are relatively light~\cite{Hook:2018jle,DiLuzio:2021pxd}. Such axions can be sourced by NSs and affect the dynamics of binary NS coalescence, leaving potentially detectable fingerprints in the inspiral waveform~\cite{Hook:2017psm,Huang:2018pbu}. To search for such axions, we perform a Bayesian analysis of GW170817 taking into account the possible dephasing caused by the axions. The posterior distribution over the waveform parameters suggests no significant evidence for such axion fields. As shown in Fig.~\ref{fig:cons}, this null result excludes a large region of the axion parameter space, much of which has not been probed by existing experiments. Importantly, our constraints are independent of the assumption that axions are the dark matter, which is required for the constraint from big bang nucleosynthesis (BBN)~\cite{Blum:2014vsa}. In this Letter, we use the conventions $\hbar=c=1$.\\

\emph{Neutron stars with axions,---}%
We consider axions that couple to nuclear matter in a similar way as the QCD axion. The low energy effective potential is \cite{Hook:2017psm}
\ba
V(a) = - m_{\pi}^2 f_{\pi}^2 \epsilon \sqrt{1 - \frac{4 m_u m_d}{(m_u+m_d)^2}\sin^2 \left(\frac{a}{2f_a}\right)}\,,
\ea
where $f_a$ is the axion decay constant, $m_{\pi}$ and $f_\pi$ are the pion mass and decay constant, and $m_{u,d}$ stands for the mass of the up, down quarks. Assuming $\epsilon < 0.1$, the mass of the axions is $m_a \simeq \sqrt{\epsilon}\, m_\pi f_\pi /f_a$, and is lighter than the mass of the QCD axion. In vacuum, the axion field is expected to stay at the minimum of its potential $a=0$. Inside a dense object, such as a NS, the axion potential receives finite density corrections \cite{Cohen:1991nk}
\ba
V(a) = - m_{\pi}^2 f_{\pi}^2 \left[\left(\epsilon - \frac{\sigma_N n_N}{m_{\pi}^2 f_{\pi}^2}\right) \left|\cos\left(\frac{a}{2f_a}\right)\right| \right.~ &&\nonumber \\
\left. + {\cal O}\left(\left(\frac{\sigma_N n_N}{m_{\pi}^2 f_{\pi}^2}\right)^2\right)\right]&&,
\ea
where $n_N$ is the number density of nucleons, and $\sigma_N \approx 59 \,{\rm MeV}$ \cite{Alarcon:2011zs}\footnote{Specific mechanisms that suppress the axion masses~\cite{Hook:2018jle,DiLuzio:2021pxd} might also change the period of this low energy effective potential. However, the axion profile and subsequent analysis is determined exclusively by the finite density effect inside the NS, with period $2\pi f_a$. Therefore, our analysis applies to the light axions in \cite{Hook:2018jle,DiLuzio:2021pxd}.
}. For $\epsilon < 1$, the axion potential inside the dense object can change sign while the perturbation theory is still valid. If the radius of the dense object is larger than the critical radius
\ba
R_{\rm crit} \equiv \frac{2 f_a}{\sqrt{\sigma_N n_N-\epsilon m_\pi^2 f_\pi^2}},
\ea
a phase transition occurs, shifting the vacuum expectation value of the axion field from $0$ to $\pm \pi f_a$ inside the dense object. Assuming NSs have a radius on the order of $10 {\rm km}$, this phase transition generally happens inside NSs for axions with $f_a \lesssim 10^{18} {\rm GeV}$. As a result, the NS develops an axion profile, interpolating from $\pm \pi f_a$ near the NS surface to $0$ at spatial infinity.

In this case, the axion field mediates an additional force between NSs, with strength that could in principle be  as strong as gravity. The axion force cannot be sourced by nuclei (as nuclei are too small to trigger the phase transition), and can therefore avoid fifth force constraints in laboratories. At leading order, the axion force between two NSs is
\ba\label{eq:force}
{\bf F}_a = - \frac{Q_{1}Q_{2}}{4 \pi r^2}\left(1+m_a r\right)\exp[- m_a r] \, \hat{{\bf r}}\, ,
\ea
where $Q_{1,2}$ is the axion charge carried by each NS and is related to the NS radius $R_{1,2}$ through
\ba\label{eq:charge}
Q_{1,2} = \pm 4 \pi^2 f_a R_{1,2}\, .
\ea
The axion force can be either attractive or repulsive, depending on whether the axion field values are of the same or opposite sign on the surfaces of the two NSs. Moreover, the axion force is only ``turned on" if the two NSs are within the axion's Compton wavelength $\lambda_a \equiv 1/m_a$.

If such NSs form binaries, the axion field might also radiate axion waves during binary coalescence. For circular orbits, the leading order radiation power is
\ba\label{eq:radiation}
P_a =  \frac{(Q_{ 1} M_2 - Q_{ 2} M_1)^2}{12 \pi \left(M_1+M_2\right)^2} r^2 \Omega^4 \left(1-\frac{m_a^2}{\Omega^2}\right)^{3/2} \, , 
\ea
where $\Omega$ is the orbital frequency and $r$ denotes the separation between the two NSs of masses $M_1$ and $M_2$. According to Eq.~\eqref{eq:radiation}, the axion radiation is turned on only when the orbital frequency is larger than the axion mass. The axion force as well as the axion radiation power are calculated to the next-to-leading order in Ref.~\cite{Huang:2018pbu}.\\

\emph{Waveform template.---}%
Inspirals in the presence of a generic massive scalar field have been studied in Refs.~\cite{Sampson:2013wia,Croon:2017zcu,Sagunski:2017nzb,Huang:2018pbu,Alexander:2018qzg,Kopp:2018jom,Seymour:2019tir,Seymour:2020yle}, among which corrections of the scalar field on the GR waveforms are calculated to the first post-Newtonian (PN) order in~\cite{Huang:2018pbu}. The waveform template cannot be written in a closed analytic form, and cannot be described by the usual PN templates, e.g.,  the one used in Ref.~\cite{Abbott:2018lct}. In our analysis, the waveform is generated by a modified TaylorF2 template, in which the frequency domain waveform is given by
\ba\label{eq:TF2}
h(f) \simeq H(f) \exp \left[ i \Psi(f) \right].
\ea
Since the existing analyses of GW170817~\cite{TheLIGOScientific:2017qsa,Abbott:2018exr} show good agreement with GR, the axion charges, if nonzero, must be very small, which allows us to expand $\Psi$ as
\ba\label{eq:Psi}
\Psi = \Psi_{\rm GR} + \Psi_a + {\cal O}(Q_{1,2}^4) + {\cal O}(Q_{1,2}^2 v^2).
\ea
Here $\Psi_{\rm GR}$ is the phase in the usual TaylorF2 template in the PyCBC package \cite{Biwer:2018osg}, $\Psi_a$ is the leading order phase correction caused by the axion field, and $v^2$ counts the PN order. The expression of $\Psi_a$ can be found in the Supplementary Material. In practice, we only consider the leading order correction caused by the axion field, which is justified by the necessary smallness of the axion charge.

Generally, taking into account the leading correction from a massive scalar field introduces three parameters in the waveform template, i.e., the scalar charge of each star and the mass of the scalar field. In our case, the two charges $Q_1$ and $Q_2$ are given by Eq.~\eqref{eq:charge} and hence are not independent. Thus, we define
\ba\label{eq:qa}
\gamma_a \equiv \frac{Q_1Q_2}{4\pi G M_1M_2},
\ea
a dimensionless parameter that characterizes the relative strength of the axion and gravitational force between the two NSs. The effects of the axion field are then parametrized by $m_a$ and $\gamma_a$. In order to obtain each charge $Q_{1,2}$ from $\gamma_a$, we first use the universal $\Lambda - C$ relation \cite{Yagi:2016bkt, Maselli:2013mva, Urbanec:2013fs} to compute the compactness and hence the radius of each NS. Then with Eq.~\eqref{eq:charge} we compute $Q_1/Q_2$, and eventually obtain the two charges $Q_1$ and $Q_2$ that are used to generate the waveform.

Moreover, we assume the two NSs obey the same equation of state (EOS), in which case their tidal deformabilities $\Lambda_1$ and $\Lambda_2$ are related. Following Ref.~\cite{Abbott:2018exr}, we consider that the symmetric tidal deformability $\Lambda_s \equiv (\Lambda_2+\Lambda_1)/2$, the antisymmetric tidal deformability $\Lambda_a \equiv (\Lambda_2-\Lambda_1)/2$ and the mass ratio of the binary $q \equiv m_2/m_1 \le 1$ are related through an EOS-insensitive relation $\Lambda_a(\Lambda_s, q)$ \cite{Yagi:2013bca,Yagi:2013awa}. In Bayesian analysis, we sample uniformly in the symmetric tidal deformability $\Lambda_s \in [0,2000]$, while $\Lambda_a$ and hence $\Lambda_1$ and $\Lambda_2$ are obtained using the EOS-insensitive relation $\Lambda_a(\Lambda_s, q)$ which is tuned to a large set of EOS models \cite{Yagi:2015pkc,Chatziioannou:2018vzf}.\\

\emph{Bayesian inference.---}%
To search for axions, we scan the parameter space by sampling axion fields with different masses (see Fig.~\ref{fig:cons} for the masses). In addition, we also consider the massless limit $m_a=0$. For each mass, we perform a Bayesian analysis of GW170817, taking into account the possible dephasing caused by the axion field in the inspiral waveform. In particular, we consider a set of parameters $\bm{\vartheta} = (\gamma_a, \bm{\vartheta}_{\rm NS})$, and evaluate the posterior probability density function $p(\bm{\vartheta}|d)$ given the GW170817 data $d$. Here $\bm{\vartheta}_{\rm NS}$ includes chirp mass ${\cal M}$, mass ratio $q$, coalescence time, coalescence phase, polarization, inclination, spins of two NSs, and symmetric tidal deformability which are defined in the usual TaylorF2 waveform template. In these analyses, we fix the luminosity distance $D_L = 40.7{\rm Mpc}$ \cite{Cantiello:2018ffy} and the sky localization $({\rm RA},\,{\rm Dec}) = (197.450374,\, -23.381495)$ \cite{Soares-Santos:2017lru} for GW170817, as they have been accurately measured independently.

In order to determine the posterior distribution over the parameters $\bm{\vartheta}$, we make use of the Markov-chain Monte Carlo algorithm as implemented in the PyCBC package \cite{Biwer:2018osg}. For the likelihood calculation, we use GW170817 data version 3 released by the LIGO and Virgo scientific collaboration on a GW open science center \cite{Abbott:2019ebz}, and assume a Gaussian noise model with a low frequency cutoff of $20~{\rm Hz}$. We only use LIGO Hanford and Livingston data, since the signal to noise ratio (SNR) of Virgo data is far smaller \cite{TheLIGOScientific:2017qsa}.

The priors on $\gamma_a$ are chosen to be $(-0.1,\, 0.1)$. The sign of $\gamma_a$ indicates whether the axion force is attractive or repulsive. In principle, the probability of an attractive or repulsive axion force can be different, depending on the formation history of the binary. Nevertheless, we assume the same prior on positive and negative $\gamma_a$ for simplicity.\\

\begin{figure*}[tp]
\centering 
\includegraphics[height=0.45\textwidth]{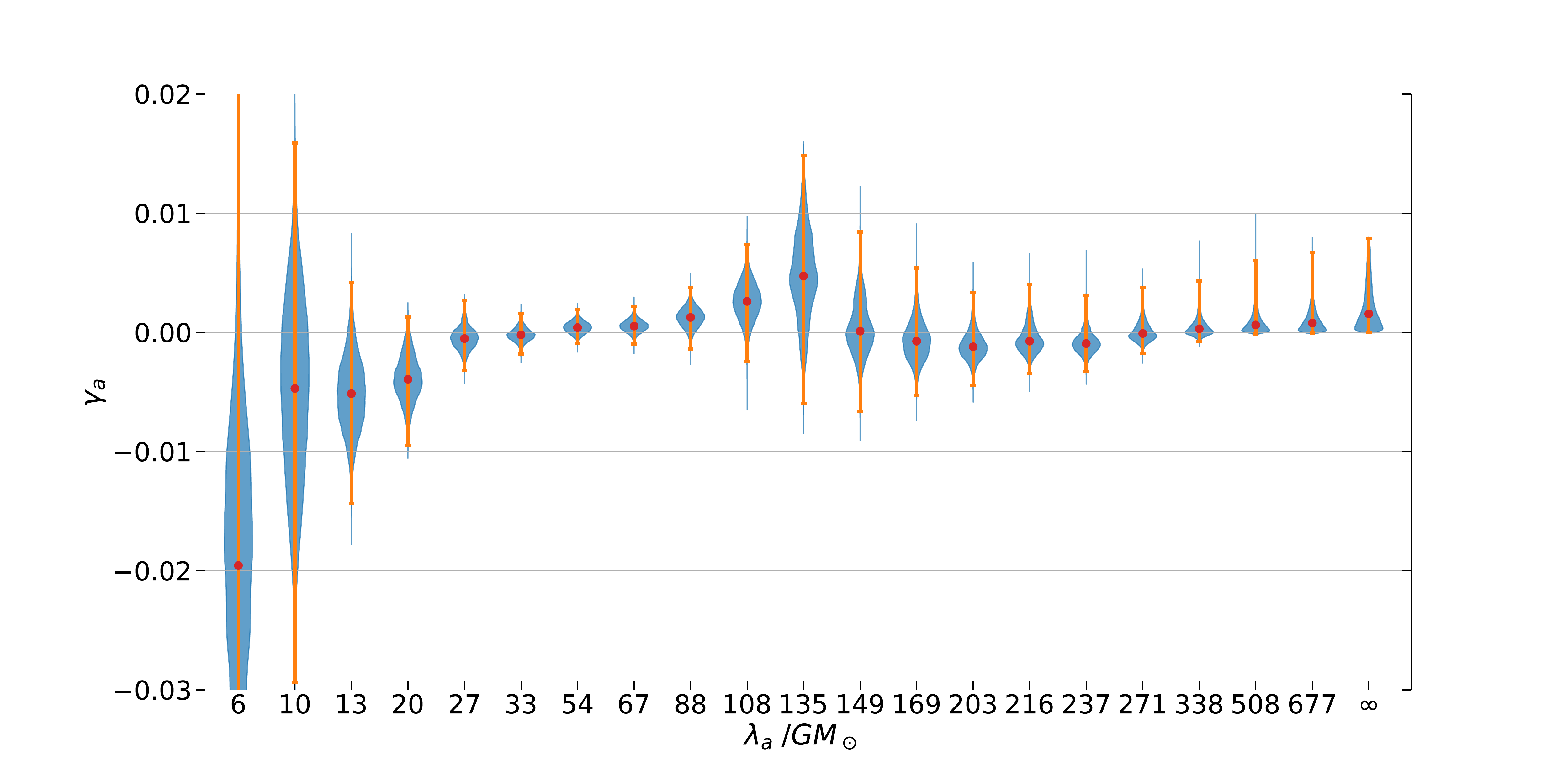} 
\caption{Posteriors on $\gamma_a$ of axions with different masses. $\lambda_a \equiv 1/m_a$ is the Compton wavelength. The horizontal bars mark the $3\sigma$ standard deviations. The deviation increases rapidly at small $\lambda_a$ as the axion effects are suppressed by $m_a$. The deviation is large around $\lambda_a \sim 135 G M_\odot$ due to the degeneracy between $\gamma_a$ and the chirp mass $\cal M$. This degeneracy is partially broken (especially for  $\gamma_a < 0$) at large $\lambda_a$ by the axion radiation. The posteriors eventually approach to that in massless limit (noted as $\lambda_a = \infty$) since the GW170817 data is insensitive to waveform with $ |\gamma_a| \le 10^{-2}$ at large $\lambda_a$.
}\label{fig:viol}
\end{figure*}

\emph{Results.---}%
We focus on the posteriors over $\gamma_a$, which are shown in Fig.~\ref{fig:viol} as a function of the axion Compton wavelength. The posteriors show no significant evidence for nonzero $\gamma_a$, and are compatible with $\gamma_a = 0$ at a $3\sigma$ confidence level over the full range of axion masses sampled.

The standard deviation of $\gamma_a$ increases dramatically when $\lambda_a$ becomes smaller than $10\, GM_\odot$, in which case $\lambda_a$ becomes less than the NS radii and the effects of the axion field is suppressed. On the other hand, for $\lambda_a \gtrsim r_{\rm cut}$, where $r_{\rm cut} \simeq 110\,G M_\odot$ is the separation between the two NSs corresponding to the $20 \rm{Hz}$ low frequency cutoff, the axion force~\eqref{eq:force} behaves like a Newtonian force during the whole observed inspiral stage. Without axion radiation, $\gamma_a$ would be highly degenerate with the chirp masses $\cal M$. This is indeed the case for $\lambda_a \sim 135 G M_\odot$, where the axion radiation is still not significant, and the standard deviation of $\gamma_a$ is large due to the degeneracy between $\gamma_a$ and $\cal M$. As $\lambda_a$ increases, the axion radiation becomes significant and breaks this degeneracy, especially for negative $\gamma_a$. For positive $\gamma_a$, the charge difference is small if the radii of the two NSs are comparable; thus, the axion radiation is always weaker than for negative $\gamma_a$. This is also why the constraints on negative $\gamma_a$ are better than those for positive $\gamma_a$ at large $\lambda_a$. The degeneracy can also be partially broken by considering the induced charge effect studied in Ref.~\cite{Huang:2018pbu}, which could improve the constraints at large $\lambda_a$ by a factor of roughly $2$.

The posteriors on $\gamma_a$ become stable for $\lambda_a >  338 G M_\odot$. This is because $|\gamma_a|$ is constrained to be smaller than $10^{-2}$ for axions with mass $\lambda_a >  338 G M_\odot$. With such a small $|\gamma_a|$, the phase difference is less than ${\cal O}(1)$, and hence  the GW170817 data has no distinguishing power. Indeed, we find that the posteriors with $\lambda_a >  338 G M_\odot$ eventually approach the posterior in the massless limit. The insensitivity of posterior on large $\lambda_a$ allows us to also impose constraints on axions with $\lambda_a >  338 G M_\odot$.

To draw conclusions about axion fields, we project the $3\sigma$ constraints on $\gamma_a$ onto the decay constant $f_a$. We combine the constraints from positive and negative $\gamma_a$ by picking the weakest one. As shown in Fig.~\ref{fig:cons}, the constraints on $\gamma_a$ translate to a constraint of $f_a < 1.6\times10^{16} {\rm GeV}$ on axions with $m_a \le 10^{-11} {\rm eV}$. On the other hand, for axions with $f_a > 10^{18} {\rm GeV}$, the critical radius is so large that NSs cannot trigger the phase transition. In other words, axions with $f_a > 10^{18} {\rm GeV}$ cannot be sourced by NSs even if they exit, and are free from the NS inspiral constraint. Therefore, our analysis indicates that GW170817 imposes constraint on axions with masses below $10^{-11} {\rm eV}$ by excluding the ones with decay constants $1.6\times10^{16} {\rm GeV} < f_a < 10^{18} {\rm GeV}$. \\

\emph{Discussion.---}%
Our analysis provides the first constraint on axions from NS inspirals, and excludes a large parameter space that has not been probed by existing experiments. As a comparison, in Fig.~\ref{fig:cons} we show the $3\sigma$ constraint from the spin measurements of the stellar mass black holes (in green)~\cite{Baryakhtar:2020gao}, as well as the $1\sigma$ constraint on axion dark matter from BBN (in red)~\cite{Blum:2014vsa}. In addition, axions can be constrained by the absence of GWs emitted by the superradiance cloud around stellar mass black holes \cite{Brito:2017wnc,Brito:2017zvb,Tsukada:2018mbp,Palomba:2019vxe,Sun:2019mqb}. Since this constraint is also based on the absence of the superradiance, it excludes a similar parameter space as the one that is excluded by the spin measurements of stellar mass black holes. We emphasize that our analysis imposes constraint on parameter space that cannot be covered by the existing experiments. For example, superradiance can only be used to probe axions whose Compton wavelength is comparable or slightly larger than the size of black holes. Therefore, the superradiance constraints, from both black hole spin measurements and the GWs emitted by superradiance clouds, cannot probe axions with very small masses due to the lack of the corresponding heavy black holes or the low superradiance efficiency. Moreover, our analysis does not rely on the assumption that the axions make up the dark matter, which is required for the BBN~\cite{Blum:2014vsa} and the neutron electric dipole moment (nEDM)~\cite{Abel:2017rtm,Du:2018uak} constraints. Especially, the kinetic energy and momentum of axions with $f_a \lesssim 10^{17} \rm{GeV}$ would change by more than ${\cal O}(1)$ near the Earth due to the finite density corrections; therefore, most of the constraint from the Earth based nEDM experiments are in question. Besides the above constraints, axions with smaller $f_a$ (in purple region in Fig.~\ref{fig:cons}) can be sourced by the Earth and the Sun for the same reason, and hence are excluded~\cite{Hook:2017psm}. Also see Refs.~\cite{Poddar:2019zoe,Poddar:2020qft} for constraints from pulsars.

We did not consider the induced charge effect, whose relative magnitude is $v^2 \exp({-m_a r})/C_{1,2}$ comparing to the axion effects considered in this Letter (see Supplementary Material). This effect could become important at the late inspirals for axions with small masses, and could potentially extend the excluded region to $10^{16} {\rm GeV} < f_a < 10^{18} {\rm GeV}$ for $m_a \lesssim 10^{-14} {\rm eV}$. However, including this effect requires further understanding on how the induced charges affect the axion radiations, and is beyond the scope of this work.

Constraint from binary NS inspirals can be further improved if the SNR of the merger event is enhanced, for example by stacking multiple binary NS merger events or with the next generation GW detectors. We expect the constraint on $f_a$ to improve by a factor of $\sqrt{N}$ if the SNR is improved by a factor of $N$. In addition, assuming a similar SNR as GW170817, the constraint could also be improved by roughly $2$ orders of magnitude if we observe a NS-black hole merger, in which case the axion radiation is not suppressed by the small charge difference and there is no degeneracy between parameters for axions with small masses. A joint analysis of the events GW190425 \cite{Abbott:2020uma} and GW190814 \cite{Abbott:2020khf}, which may contain NSs, is left for future work.\\

{\it Acknowledgements---} 
We thank Katerina Chatziioannou, Claudia de Rham, and Nicol\'as Yunes for discussions. J.Z. is supported by the European Union's Horizon 2020 Research Council grant 724659 MassiveCosmo ERC-2016-COG. M.C.J. and H.Y. are supported by the National Science and Engineering Research Council through a Discovery grant. M.S. is supported in part by the Science and Technology Facility Council (STFC), United Kingdom, under research grant No. ST/P000258/1. This research was supported in part by Perimeter Institute for Theoretical Physics. Research at Perimeter Institute is supported by the Government of Canada through the Department of Innovation, Science and Economic Development Canada and by the Province of Ontario through the Ministry of Research, Innovation and Science. This is LIGO document number LIGO-P2100161.

\bibliography{references}

\appendix

\section{Supplementary Material}

In the presence of the axions, the leading corrections to the GR binding energy and radiation power are given by
\ba
V_a = -\frac{Q_1Q_2}{4\pi} \frac{e^{-m_a r}}{r}\left(1-16 p\,GM \frac{e^{-m_a r}}{r}\right)
\ea
and
\ba
P_a = \frac{\left(\bar{Q}_1M_2-\bar{Q}_2M_1\right)^2}{12\pi} r^2 \Omega^4 \left(1-\frac{m_a^2}{\Omega^2}\right)^{3/2}
\ea
with
\ba
p \equiv \frac{1}{M}\left(\frac{Q_1}{Q_2} p_2 + \frac{Q_2}{Q_1}p_1\right)\,
\ea
and
\ba
\bar{Q}_{1,2} \equiv Q_{1,2} \left(1 -8Gp_{2,1} \frac{e^{-m_a r}}{r} \right)\,.
\ea
Here $\Omega$ is the orbital frequency, $r$ denotes the separation between the two NSs of masses $M_1$ and $M_2$, and $M \equiv M_1+M_2$ is the total mass. Comparing to Eqs.~\eqref{eq:force} and \eqref{eq:radiation}, we also include terms proportional to $p_{1,2}$ that could in principle arise due to the present of a generic scalar field. The value of $p_{1,2}$ is model dependent. For axions, these terms characterize the induced charge effect, and $p = \left(R_1 + R_2\right)/16GM$ when $m_a = 0$. Thus, we expect that the induced charge effect could become important at the late stage of inspirals for axions with small masses. However, taking into account this effect requires further studies on how $p_{1,2}$ relates to the parameters of the neutron stars and the axion field. Therefore, we neglected the induced charge effect in our analysis.

In TaylorF2 waveform temple, the phase $\Psi(f)$ in Eq.~\eqref{eq:TF2} can be calculated by using the stationary phase approximation, 
\ba
\Psi(f) =  2\pi f t-\phi -\frac{\pi}{4}
\ea
with
\ba\label{tf}
t(f)  = t_c - \int_{f_c}^{f} \frac{1}{P} \left(\frac{dE}{df'}\right)df'
\ea
and
\ba\label{phif}
\phi(f)  =\phi_c - \int_{f_c}^{f} \frac{2\pi f' }{P} \left(\frac{dE}{df'}\right)df',
\ea
where $E$ and $P$ are the binding energy and radiation power of the binary system respectively. Note that $E$ and $P$ are functions of $r$ and $\Omega$, which are related by the modified Kepler's law $r(\Omega)$, and the GW frequency relates to the orbital frequency through $\Omega = \pi f$. Given the fact that the axion charge $Q_{1,2}$ must be small, we neglect terms of ${\cal O}(Q_{1,2}^4)$, ${\cal O}(Q_{1,2}^2 v^2)$ and higher when we evaluate Eqs.~\eqref{tf} and~\eqref{phif}. In this case, the phase $\Psi(f)$ is given by Eq.~\eqref{eq:Psi} with
\ba
\Psi_a  = \Psi_a^E + 
\begin{cases}
\Psi_a^{P_>} &  x> \alpha\\
\Psi_a^{P_<} & x \le \alpha
\end{cases}\,,
\ea
where we have defined
\begin{widetext}
\ba
\Psi_a^E &=& \frac{5}{64} \frac{\gamma _a e^{-\frac{\alpha }{x^{2/3}}}}{\eta  x^{5/3}}  \left[-4-\frac{32 x^{2/3}}{\alpha } -\frac{138 x^{4/3}}{\alpha ^2}-\frac{360 x^2}{\alpha ^3}+\frac{360 x^{8/3} \left(e^{\frac{\alpha }{x^{2/3}}}-1\right)}{\alpha ^4}-\frac{21 \sqrt{\pi } x^{5/3} e^{\frac{\alpha }{x^{2/3}}} \text{Erf}\left(\frac{\alpha^{1/2}}{x^{1/3}}\right)}{\alpha ^{5/2}}\right],\\
 \Psi_a^{P_>} &=&\frac{5 }{254951424} \frac{\delta q^2}{\eta  x^{16/3}} \left [ -\sqrt{x^2-\alpha ^2} \left(-822640 \alpha ^2+\frac{227089 x^6}{\alpha ^4}+\frac{261342 x^4}{\alpha ^2}+671304 x^2\right) \right. \nonumber \\
&&+\frac{140049 x^7}{\alpha ^4}\, {}_2F_1\left(-\frac{5}{6},\frac{1}{2};\frac{1}{6};\frac{\alpha ^2}{x^2}\right) + 320 x \left(1183 \alpha ^2+\frac{512 x^6}{\alpha ^4}-\frac{684 x^4}{\alpha ^2}-741 x^2\right) \, _2F_1\left(-\frac{1}{2},-\frac{1}{3};\frac{2}{3};\frac{\alpha ^2}{x^2}\right)  \nonumber \\
&&+\left.960x \left(-1183 \alpha ^2-\frac{80 x^6}{\alpha ^4}+\frac{684 x^4}{\alpha ^2}+741 x^2\right) \, _2F_1\left(-\frac{1}{3},\frac{1}{2};\frac{2}{3};\frac{\alpha ^2}{x^2}\right)\right] \, , \\
\Psi_a^{P_<} &=&\frac{25\sqrt{\pi}}{1536} \frac{{\delta q}^2}{\alpha ^{10/3} \eta} \frac{\Gamma \left(\frac{5}{3}\right) \Gamma \left(\frac{11}{3}\right) x -  \Gamma \left(\frac{7}{6}\right) \Gamma \left(\frac{25}{6}\right)\alpha}{   \Gamma \left(\frac{11}{3}\right) \Gamma \left(\frac{25}{6}\right)} \, .
\ea
\end{widetext}
Here $\text{Erf}$ is the Gauss error function, $_2F_1$ is the hypergeometric function, and
\ba
&&\alpha \equiv GM m_a, \quad \eta \equiv \frac{M_1M_2}{M^2}, \quad x \equiv \pi GMf\, , \nonumber \\
&&\delta q \equiv \frac{1}{4\sqrt{2\pi G} }\left(\frac{Q_1}{M_1}-\frac{Q_2}{M_2}\right).
\ea

\end{document}